# Considerations for Master Protocols Using External Controls

Jie Chen[1]*, PhD, Xiaoyun (Nicole) Li[2], PhD, Chengxing (Cindy) Lu[3], PhD,
Sammy Yuan[4], PhD, Godwin Yung[5], PhD, Jingjing Ye[6], PhD, Hong Tian[7], PhD,
Jianchang Lin[8], PhD

[1]ECR Global, Shanghai, China

[2]BeiGene, San Mateo, CA

[3]AstraZeneca, Boston, MA

[4]GlaxoSmithKline, Collegeville, PA

[5]Roche, San Francisco, CA

[6]BeiGene, Fulton, MD

[7]BeiGene, Township, NJ

[8]Takeda, Cambridge, MA

---

*Correspondence to: Dr. Jie Chen, jiechen0713@gmail.com, ECR Global, No. 391 Huiping Road, Shanghai, China

# Considerations for Master Protocols Using External Controls

**Abstract**

There has been an increasing use of master protocols in oncology clinical trials because of its efficiency and flexibility to accelerate cancer drug development. Depending on the study objective and design, a master protocol trial can be a basket trial, an umbrella trial, a platform trial, or any other form of trials in which multiple investigational products and/or subpopulations are studied under a single protocol. Master protocols can use external data and evidence (e.g., external controls) for treatment effect estimation, which can further improve efficiency of master protocol trials. This paper provides an overview of different types of external controls and their unique features when used in master protocols. Some key considerations in master protocols with external controls are discussed including construction of estimands, assessment of fit-for-use real-world data, and considerations for different types of master protocols. Similarities and differences between regular randomized controlled trials and master protocols when using external controls are discussed. A targeted learning-based causal roadmap is presented which constitutes three key steps: (1) define a target statistical estimand that aligns with the causal estimand for the study objective, (2) use an efficient estimator to estimate the target statistical estimand and its uncertainty, and (3) evaluate the impact of causal assumptions on the study conclusion by performing sensitivity analyses. Two illustrative examples for master protocols using external controls are discussed for their merits and possible improvement in causal effect estimation.

## 1   Introduction

A master protocol refers to a framework in which multiple substudies are conducted under the same unifying ("master") protocol and substudies are differentiated by either multiple investigational products (IPs) or multiple diseases or disease subtypes (Woodcock and LaVange, 2017; Renfro and Mandrekar, 2018; Bogin, 2020). The US Food and Drug Administration (FDA) guidance defines a master protocol as "a protocol designed with multiple substudies, which may have different objectives and involve coordinated efforts to evaluate one or more investigational drugs in one or more disease subtypes within the overall trial structure" (FDA, 2022). Depending on the design features, e.g., simultaneous evaluation of multiple IPs, multiple diseases or disease subtypes, or both, a master protocol trial can be a basket trial, an umbrella trial, a platform trial, or any other form of trials in which multiple IPs and/or multiple subpopulations are investigated, either in parallel o sequentially, under a single protocol to achieve better operational efficiency. A master protocol is called a *basket trial* if it is designed to evaluate a single IP or drug combination in different (sub)populations defined by different diseases or disease subtypes. On the contrary, a master protocol is called an *umbrella trial* if it is designed to evaluate multiple IPs, each of which is administered as a single agent or as a drug combination in a single disease population. A master protocol is called a *platform trial* if multiple IPs and/or drug combinations are investigated in a perpetual manner (i.e., an IP can leave the trial if it doesn't show efficacy and new IPs can enter over the course of the trial) across multiple diseases or disease subtypes (Woodcock and LaVange, 2017; Lu et al., 2021).

Some design features in master protocols include: (1) substudies in basket trials are usually designed as single-arm trials with each substudy having specific clinical questions to be addressed, scientific rationale, study objectives, and statistical analysis plan, and (2) umbrella trials often use a common control arm, e.g., standard of care (SoC), to compare multiple IPs simultaneously in the same target population. In basket trials, an external control is used, either explicitly or implicitly, in comparative analyses for each substudy; in umbrella trials, the common control is often an internal concurrent or non-concurrent control, which can also be augmented by external control data, the so-called *hybrid control*, e.g., Desai et al. (2019).

In this paper external data refer to as data generated from outside the planned master protocol trial and external evidence is evidence derived from such defined external data. External data may include data from completed clinical trials and real-world data (RWD) (FDA, 2023). The US FDA defines the real-world data (RWD) as "data relating to patient health status and/or the delivery of health care routinely collected from a variety of sources" and real-world evidence (RWE) as "the clinical evidence about the usage and potential benefits or risks of a medical product derived from analysis of RWD" (FDA, 2018). There are basically two types of RWD—research-oriented RWD and transaction-oriented RWD (Franklin and Schneeweiss, 2017). Examples of research-oriented RWD are data from disease registries (e.g., the Surveillance, Epidemiology, and End Results [SEER] registries), product registries (focusing on patients treated with a particular products, often used for safety and real-world effectiveness evaluation of an approved product), RWE studies (e.g., real-world clinical studies aiming to generate real-world effectiveness and safety), and longitudinal cohort follow-up studies (e.g., Framingham study). Examples of transaction RWD include electronic health/medical records (EH/MRs), claims data, and pharmacy dispensing data. It is generally believed that research-oriented RWD tend to be more structured for research purpose than transaction data and hence in better quality with respect to accuracy, completeness, consistency, and reliability (Levenson et al., 2023; Chen et al., 2023).

The purpose of using external data and evidence (EDE) in clinical trials (including master protocols) can largely be classified into three categories: (1) informing trial design (e.g., providing information on clinical characteristics of a disease, patient backgrounds and size, SoC, selection of endpoints, etc.), (2) assisting trial conduct (e.g., site selection, patient recruitment and retention), and (3) serving as external controls (either as a fixed value derived from external data or as an external control arm). While EDE has traditionally been used in the first two categories (see, e.g., Friedman et al. (2015), Mayo et al. (2017), Chen et al. (2023), Rahman et al. (2021), and Chen et al. (2023) for more discussion), use of external controls in category 3 has recently received increasing attention (see, e.g., Lim et al. (2019); Ghadessi et al. (2020); Jahanshahi et al. (2021); FDA (2023), and references therein) due to, e.g., considerations of infeasibility or ethical concerns of randomized controlled trials (RCTs). For example, for diseases with an extremely low incidence/prevalence and/or unmet medical needs, RCTs may not be practical or even unethical, in which case external controls may be used for treatment effect estimation. However, due to the mostly untestable causal assumptions (see Section 4.5), estimation of treatment effects is much more complex. On the other hand, there is scarce literature discussing the use of external controls in master protocols, which prompts the American Statistical Association (ASA) Biopharmaceutical Section (BIOP) Statistical Methods in Oncology Scientific Working Group (SWG) Master Protocol Subteam to develop this paper to provide some key considerations when using external controls in master protocols including construction of estimands, assessment of for-for-use external data, and a roadmap for using EDE in master protocols (either exploratory or confirmatory) for drug development.

The rest of the paper is organized as follows. Section 2 describes different types of exter-nal controls and associated technical considerations when using them in master protocols. Section 3 expands the discussion on general considerations described in FDA (2023) with a special focus on construction of estimands including assessing fit-for-use external data as external controls in master protocols. Section 4 presents a causal roadmap when using EDE in master protocols and Section 5 provides two case studies. Finally, some discussions and concluding remarks are given in Section 6.

## 2  External Controls for Master Protocols

ICH E10 (ICH, 2001) states that an external controlled trial compares "a group of subjects receiving the test treatment with a group of patients external to the study, rather than to an internal control group consisting of patients from the same population assigned to a different treatment. The external control can be a group of patients treated at an earlier time (historical control) or a group of patients treated during the same time period but in another setting [contemporaneous external control]." The US FDA draft guidance on rare disease (FDA, 2019) points out that use of external controls restricts to clinical trials in the following scenarios: (1) life-threatening diseases with unmet medical needs, (2) diseases with well-understood and highly predictable natural history, and (3) an expected substantial drug effect that is self-evident and temporally closely related to the treatment. These three conditions are perceivably applicable to some master protocols.

Depending on temporality of the control data generation relative to the start of a master protocol, control group data can be classified as (1) retrospective (historically collected data), (2) prospective (prospectively collected data), and (3) retro-prospective (including both historically and prospectively collected data, or historical-contemporaneous data). Note that prospective data are always preferred over retrospective data since the former may have better quality through prospectively designed algorithms for data collection. Control data can be either patient-level data or study-level summary data with data sources including completed clinical trials, RWD (e.g., EH/MRs, disease registries), and literature reports. Retrospectively collected data can be used in a *historical control* or a *historical-contemporaneous control*, whereas prospectively collected data can be used in a contemporaneous external control or simply *contemporaneous control* if the control subjects are not part of the study subjects. The internal control can also be either a *concurrent control* if the data for both control and treatment groups are collected in the same time period or a *non-concurrent control* if the data for both control and treatment groups are collected in different time periods. In some cases, external control data may be combined with internal concurrent control data to form a *hybrid control* arm (see Section 2.6). Often a data-combination algorithm is used to pool multiple control data sources to create a *synthetic control* (see Section 2.5). Figure 1 illustrates different control types and data sources that are typically encountered in master protocols. The rest of this section describes in more details each of the control types and some special considerations when using them in master protocols.

## 2.1 Historical control

Historical control (HC) is perhaps one of the most commonly used external controls in clinical research as it conveniently uses existing (historically collected) data for estimation of treatment effect. There is rich literature discussing how historical controls can be used in drug development and regulatory decision-making, e.g., Viele et al. (2014), Lim et al. (2018), Lim et al. (2019), Ghadessi et al. (2020), Jahanshahi et al. (2021), and references therein. Although HC data may come from multiple sources such as those described in Section 1, the most suitable data sources of HC for master protocols are perhaps disease registries (e.g., Okuma et al. (2020)) and some research-oriented databases, (e.g., Chau et al., 2018; Oh et al., 2020; Ko et al., 2020), especially for rare diseases and diseases with unmet medical needs because this type of data sources contains rich information on patient characteristics and natural history of the diseases (as compared to transaction data).

In addition to common requirements for external controls (e.g., subjects eligibility, disease specification and diagnosis, data quality, variable ascertainment, etc.) (Pocock, 1976; Jahanshahi et al., 2021; Mishra-Kalyani et al., 2022), the following special considerations should be taken when using HCs in master protocols. First, the index window, the time period during which the first and last patients took the control product, should be chosen by considering the following: (1) clinical practice in specific disease areas—if the landscape of clinical practice (e.g., therapeutic landscape and/or disease diagnostic tools and criteria) changes rapidly, then the index window should be narrower and not too far away from the start date of the master protocol; (2) availability of the control product(s)—all the control products should be available for patients within the index window; (3) control sample size—the larger the sample size required, the longer the index window is needed; and (4) medical coding—preferably no major change in medical coding (such as MedDRA, WHODrug, etc) occurs within the index window. The FDA guidance on externally controlled trials (FDA, 2023) points out that, upon the index window appropriately determined, eligible patients should be included in the analysis regardless of survival to avoid *immortal bias* that can be introduced when eligibility determination and treatment initiation are not synchronized. Note that immortal bias can occur in any externally (not just historically) controlled studies if there is an immortal time period during which the outcome of interest could not be observed in one of the comparison arms.

Second, multiple data sources are often needed to have integrated information of control patients. For example, the primary source of control patients is a disease registry that may contain incomplete genomic information, which prompts the use of separate -omics data to be integrated with the registry data for biomarker-driven master protocols. Third, some historical data may reliably record only clinical outcomes (e.g., survival), not surrogate (e.g., tumor responses) or intermediate outcomes (e.g., progress-free survival), which may hamper the use of historical control in master protocols as the latter often uses surrogate or intermediate endpoints for expeditious treatment decision. Fourth, multiple historical controls can be used to either create a synthetic historical control (see Section 2.5) or serve as separate historical control arms to ensure consistency in treatment effect estimation, e.g., Kraus et al. (2022). See Ghadessi et al. (2020) for more discussions in minimizing limitations of using HC in clinical trials in general.

## 2.2 Contemporaneous control

Contemporaneous control (CC) refers to a control cohort of subjects (1) who are not part of the master protocol, (2) who receive control treatment (e.g., SoC) during the same time period, and (3) whose (baseline and follow-up) data are collected during the same time as the data collected for the subjects in the master protocol. Unlike in HC discussed in Section 2.1, subjects in CC are usually chosen from disease registries and/or EH/MRs who are diagnosed with the same health condition but receive real-world medical care (often an SoC) during the same time as those in the master protocol. Some special considerations should be taken when using CC in master protocols. First, disease diagnosis tools may differ between the master protocol and real-world practice, although the diagnosis criteria might be the same, making it difficult to compare subjects as they may be different populations. For example, disease progression may be confirmed using both clinical evidence and radiologic evidence in a master protocol, but only one of these two pieces of evidence in CC. Second, disease progression may be assessed differently among subjects in the master protocol and in the CC, e.g., the former may use an independent review based on both clinical and radiological evidence with stringent response assessment criteria, while the latter may use clinical judgement of treating physicians. Third, more diverse treatment patterns and intercurrent events may be expected in CC than in the master protocol, as will be discussed in Section 3.1.

## 2.3 Non-concurrent control

Non-concurrent control (NCC) is an internal control comprising a group of control subjects who are part of the master protocol receiving the designated control drug but whose treatment, follow-up, and outcome collection period differ from that for the treatments to be compared in the master protocol (see Figure 1). Although not an external control, NCC is quite common in platform trials in which NCC is initiated at a time point that is different from that for the treatment arms. Because of non-concurrence in nature, subjects are not randomized head-to-head, leading to possible imbalanced distribution of baseline covariates; in addition, unblinding to treatment and outcome assessment could be problematic. Therefore, NCC may cause potential bias and type I error inflation. In addition, in comparison with any external controls (e.g., CC or HC), NCC has the advantages of using the same infrastructure of the master protocol (e.g., the same outcome assessment and data collection methods), but may suffer from time drifts in patient population, medical practice, and SoC. To account for this time trend, one may consider using the Bayesian time machine to model potential temporal drift in the entire study population for time-adjusted analysis (Saville et al., 2022). See Sridhara et al. (2022) for more discussion on the use of NCC in master protocols in oncology trials.

## 2.4 Historical-contemporaneous control

Sometimes HC data contain insufficient number of control subjects, which may prompt the use of subjects from both historical data and contemporaneous data. Major considerations in using historical-contemporaneous control (HCC) include those as discussed in Sections 2.1 and 2.2. In addition, some further considerations for master protocols with HCC may include (1) subject heterogeneity with respect to prognostic factors, (2) potential treatment heterogeneity bias (see Section 4 for more discussion), and (3) different outcome assessment methods among subjects in HC, CC, and master protocol.

## 2.5 Synthetic control

If multiple data sources are chosen to form an external control, a synthetic method can be used to create a synthetic control (SC) arm in which selected control data are assigned different weights to reflect the relative contribution of each data source to the *counterfactual* of interest and the weights are usually restricted to be positive and sum up to one (Abadie et al., 2010); see also Thorlund et al. (2020) for more discussion on synthetic controls for clinical trials in general.

Synthetic controls are particularly useful in master protocols for rare diseases because eligible population pool in each of available data sources could be small and simply pooling the control subjects from different sources may not be ideal. To create an appropriate SC, one may consider the following key steps: (1) assess fit-for-use external data (see Section 3.2) to ensure that all data sources are relevant to answer the clinical question of interest, (2) evaluate selected external data for their similarities and differences in some important aspects such as data standards, disease diagnostic tools and criteria, coding methods for diseases and drugs, completeness of subject information (especially treatment, outcome, and key covariates), etc., (3) choose a synthetic method to create an SC that best matches the treatment group in the master protocol, and (4) perform sensitivity analyses to investigate the impact of different synthetic methods, in addition to potential violation of different causal assumptions as discussed in Section 4, on the estimated treatment effect and study conclusion. See Abadie (2021) for feasibility check, data requirements, and methodological discussion, Firpo and Possebom (2018) for effect estimation, sensitivity analysis, and uncertainty assessment of synthetic control methods, and Thorlund et al. (2020) for practical considerations when using synthetic controls in clinical research.

## 2.6 Hybrid control

A hybrid control can be created by combining external control data and internal concurrent control data using a pre-defined algorithm, e.g., Bayesian borrowing (Spiegelhalter et al., 2004) and test-and-pool (Viele et al., 2014). The hybrid control method is also called internal control augmentation or dynamic borrowing in literature (Chen et al., 2023; Freidlin and Korn, 2023). The idea of hybrid control originates from Stuart and Rubin (2008) who propose an algorithm to obtain matches from multiple control groups using propensity score (PS) approach. The method can be extended to RCTs in which the internal concurrent control (CCC) (primary) cannot provide enough matches and additional (secondary) controls are needed to match the unmatched treated subjects. Yuan et al. (2019) argue that PS may not be needed for matching the CCC with the treated subjects when randomization is used to balance the covariates and hence propose a modified version that may potentially improve study power. In their approach, a randomly selected subset of treatment group is matched with CCC without using PS and the remaining unmatched subjects in treatment group are matched with external controls by propensity scores.

A hybrid control can be particularly useful in master protocols for rare diseases or disease types for which the internal control group is generally small and an augmentation can improve study power and precision of treatment effect estimation. A hybrid control for augmentation of internal control can also be considered if the compound under investigation shows an encouraging benefit-risk profile that supports an RCT with relatively smaller control group so that more subjects can benefit from the new treatment. An advantage of hybrid control is that one can assess the agreement between the internal control and external control and decide the amount of external control data to be borrowed, i.e., borrowing more if a high-level agreement is observed and less otherwise (Viele et al., 2014).

## 2.7 Virtual control

Construction of a virtual control involves developing a prediction model based on external data of treatment-naive patients and predicting the *counterfactual outcomes* by plugging the data of treated patients in the master protocol into the prediction model. Such predicted outcomes can be viewed as *virtual* outcomes to be compared with the *observed* outcomes from the same set of patients receiving the treatment of interest (Jia et al., 2014; Strayhorn, 2021). Virtual controls can be very helpful in master protocols because separate internal controls are often unavailable and there exist unmet medical needs in substudies.

The virtual control method relies on the construction of precise and accurate prediction models using treatment-naive patient baseline and time-varying covariates (including prognostic factors and clinical characteristics). Multiple prediction models can be developed based on fit-for-use external data that can be divided into training and validation datasets, to ensure acceptable performance and generalizability of predictive results (Jia et al., 2014; Switchenko et al., 2019). One may also consider using machine learning and super learning approaches of van der Laan et al. (2007) to achieve optimal prediction of virtual outcomes.

# 3 Considerations for Master Protocols Using External Controls

The FDA guidance on externally controlled trials (FDA, 2023) provides detailed discussions on considerations regarding study design, selection of external control arm, and data analysis. In particular, the guidance discusses (1) the estimand framework that involves a precise description of elements of an estimand as determined by the study objective to help design externally controlled trials and (2) data considerations for external control arm. In the rest of this section, we expand the discussion on considerations for specification of estimand attributes and assessment of fit-for-purpose external data for master protocols using external controls. We also provide special considerations for different types of master protocols (e.g., basket trials, umbrella trials, and platform trials) and summarize the similarities and differences between master protocol and non-master protocol trials when using external controls.

## 3.1 Estimands in master protocols using EDE

One of the essential steps in clinical research including master protocols is to clearly state the clinical question of interest which determines the estimand, the quantity to be estimated in the study. The International Council for Harmonisation of Technical Requirements for Pharmaceuticals for Human Use (ICH) E9(R1) guidance (ICH, 2021) presents principles for constructing estimands in clinical trials with a focus on five attributes—population, treatment, endpoints, intercurrent events, and population-level summary. Chen et al. (2023) provide considerations for constructing estimands and a roadmap in choosing appropriate estimands for RWE studies. Unlike traditional clinical trials (e.g., a randomized controlled trial or a single-arm trial) where the primary objective is to answer a single clinical question (e.g., whether a new therapy can prolong the survival of patients with a pre-defined cancer type), a master protocol often comprises multiple substudies, each of which has its own clinical questions and objectives, therefore leading to multiple estimands in a master protocol (exceptions are discussed in Section 6). The rest of this section discusses considerations in describing each of the five attributes of an estimand when using EDE in master protocols. Key covariates are also discussed as they are important in retuning the estimands in a iterative process to ensure that an reliable estimate can be achieved.

*Population.* A master protocol may have multiple target populations such as patient populations identified by different tumor types in a basket trial or a platform trial, each of these populations representing a single disease or indication. The eligibility criteria described in the master protocol define the target populations for each substudy. However, it should be noted that patients enrolled in a master protocol may not be representative of the target patient populations who would most likely use the product upon approval, not only because there are relatively restrictive inclusion and exclusion criteria for the master protocol, but also because there are (1) patients who are eligible for but choose not to participate in the study or (2) under-represented patients populations (e.g., ethnic minorities) who are eligible for, but otherwise do not have access to facilities of the study (Unger et al., 2019).

Note that when the control patients are chosen from external data (especially RWD), the patient populations in the control group may be (1) more heterogeneous than those in the master protocol with respect to demographic backgrounds, clinical characteristics, and socioeconomic status and (2) less representative of the target population due to, e.g., healthy worker effect for patients in commercial insurance claims data (Chen et al., 2023).

*Treatment.* Treatment options for patients in a master protocol are pre-defined and hence relatively fixed. For example, patients with different tumor types may constitute individual cohorts receiving a single molecularly targeted therapy in a basket trial or patients with a single tumor type can be assigned to different groups receiving respective therapies in an umbrella trial, although some rescue therapies can be administered if the assigned treatment fails. However, treatment patterns for patients in the external control group can be very complex due to, e.g., availability of multiple treatment options, personal preference of one treatment over others, etc., which reflects dynamic treatment regimes in the real-world medical practice. In addition, patients in the external control may have multiple conditions, leading to the use of concomitant medications.

*Endpoints.* Master protocols often use surrogate endpoints such as tumor response as primary endpoints in oncology study in order to expedite drug development. Such endpoints in master protocols are generally obtained based on clinical and radiologic evidence at pre-scheduled time points after treatment initiation and are often adjudicated by a third party, e.g., independent review committee. On the contrary, surrogate endpoints in external data (especially RWD) can be obtained based on either clinical evidence or radiologic evidence (but may not both) and are usually not adjudicated by a third party. In addition, surrogate endpoints in RWD may only be measured and recorded when patients visit treating physicians who observe disease progression or prescribe the radiologic examination. Therefore, missing mechanism and pattern of outcome measures in RWD may be quite different from those in master protocol; see Rockhold and Goldstein (2020) for more discussion on informed presence and non-presence of data points.

*Intercurrent events.* Intercurrent events (ICEs) in master protocols could be induced by IPs, such as events due to intolerability and/or lack of efficacy. There are also some terminal events that affect the existence and/or measurement of the endpoints when they are not part of endpoint-defining attributes. In routine clinical practice, some ICEs like treatment switching may be related to patient behavioral factors such as patient preference, recommendations by a friend or family member, or patient-physician relationship, and some other ICEs may be caused by non-behavioral factors such as change of health insurance plan (e.g., the new plan doesn't cover the current therapy) or participating in a clinical trial that requires discontinuation of the current therapy. Chen et al. (2023) propose five categories of ICEs that may potentially affect the specification and subsequent estimation of treatment effect and suggest precise definitions of all plausible ICEs in constructing an estimand in RWE studies. Note that different patterns in the occurrence of ICEs between treatment arm in a master protocol and external control arm may cause biased estimates of treatment effect if the dynamic imbalanced patient characteristics are not taken into account.

*Population-level summary and sensitivity analysis.* The choice of population-level summary will depend on the type of endpoints and the clinical question. For example, a master protocol may use the difference in tumor response rate between treatment arm and external control arm as a population-level summary in causal inference. The identification (i.e., precisely known if the population were infinite) of this causal estimand from the observed data relies on identifiability assumption (see Section 4.5). In master protocols with external controls, such an assumption may not hold. Consequently, sensitivity analyses should be performed to evaluate the impact of violation of causal assumptions on the robustness of inference. See Section 4.7 for more discussion on sensitivity analysis.

*Covariates.* Covariates are not among the five attributes of estimands in ICH (2021). However, the assumptions necessary for identifying all estimands in non-randomized studies (including master protocols using EDE) rely critically on them. The essential assumptions for identifiability of a causal estimand is conditional independence based on a set of measured covariates. Chen et al. (2023) point out that when the estimand involves a point treatment, the covariates are typically measured at baseline; when the estimand involves a time-varying treatment, the covariates may also be time-varying. Some additional considerations may be helpful when using EDE in master protocols. First, baseline covariates are often measured during a baseline window (e.g., 12 months prior to index date when treatment starts), which can lead to ascertainment bias (Anes et al., 2021). Second, some covariates such as laboratory test results may not be captured in the control patients, leading to unknown mechanisms of missingness. Third, some covariates may be measured in different scales by different methods with unknown validation parameters. See Chen et al. (2023) and Levenson et al. (2023) for more discussion on covariate ascertainment.

## 3.2 Assessment of fit-for-use external data

Given a wide range of external data sources as shown in Figure 1, an imminent task is to determine which data source(s) is fit-for-use to serve as an external control for a master protocol. The FDA RWE framework (FDA, 2018) outlines the principles of assessing fitness of RWD for regulatory decisions, which encompasses reliability and relevance of the underlying data. Reliability includes data accrual and data quality control (e.g., data completeness, consistency, trends over time, and reporting standards) and relevance concerns whether the data capture critical data elements such as exposure, outcomes, and key covariates that can be used to address the clinical and/or regulatory question, in part or as a whole.

The white paper by Duke-Margolis Center for Health Policy on determining RWD's fitness for use proposes a minimum set of standardized verification checks for completeness, conformance, and plausibility for assessing reliability across all data sources, while emphasizing the need for transparency of data provenance, accrual, and curation during the assessment (Mahendraratnam et al., 2019). Assessment of reliability and relevance of RWD involves ascertainment and validation of exposure, outcome, and key variables. The landscape assessment paper on data fitness by the ASA BIOP RWE SWG (Levenson et al., 2023) summarizes the principles for and approaches to ascertaining key variables extracted from RWD for phenotyping patients, i.e., identifying patients with certain clinical characteristics of interest using rule-based methods with clear diagnosis and procedure codes in structured data elements (Banda et al., 2018), or machine learning methods with some algorithms to classify patients with the phenotype of interest based on structured or unstructured data elements (Gibson et al., 2021). Lack or imperfect capture of some critical information in external data may result in multiple biases that threaten the fit-for-use external data sources. Levenson et al. (2023) also present the principles for and approaches to validating outcomes and assessing bias using RWD. Specifically, they discuss information bias due to misclassification and systematic measurement error, which can be categorized as non-differential error that occurs equally between groups and differential error that occurs unequally between groups, with the former driving the effect estimate toward the null and the latter driving the direction of bias (i.e., either over- or under-estimation of the effect). Therefore, outcome measures including measurement schedule, methods, and missing values should be sufficiently similar between patients in the master protocol and those in the external control.

Some data-driven approaches have been developed to evaluate various aspects of fit-for-use RWD sources. Kalincik et al. (2017) propose a quantitative approach to quantifying data completeness, accuracy and consistency as measured by data density score, error rate, and generalizability score. The data density score assesses data completeness across appropriate domains such as patient demographics, clinical characteristics, follow-up, clinical visits, and history of therapies; the error rate measures syntactic accuracy (e.g., the proportion of records falling within the realistic range of a given variable) and consistency (e.g., counting incomplete or erroneous dates); the generalizability score evaluates data believability defined as the proportion of data entries that are regarded credible with respect to the disease epidemiology (e.g., incidence, prevalence, treatment, and burden of the underlying disease). Levenson et al. (2023) develop three-dimension metrics to evaluate data fitness: (1) *Relevance based on a specific research question*: (a) disease population—number and percentage of subjects meeting the disease definition, (b) outcome—number and percentage of subjects meeting the response criteria, (c) exposure—number and percentage of subjects in treatment (control) group meeting the treatment (control) definition, (d) confounders—percentage of subjects with information on key confounding variables, (e) time—variables on time duration from study entry to response or censoring, (f) generalizability score—representativeness of the disease specific population in the data source. (2) *Reliability based on a specific research question*: (a) quality—a summary measure expressed as the percentage of data points being in error for identified key variables, and (b) completeness—a summary measure expressed as the amount of missing data for each key variable, accounting for data points that are truly missing. (3) *Fit-for-research*: (a) data provenance and traceability—qualitatively describing data provenance and manipulation as well as data traceability after curation, and (b) data formatting and standard—describing data format changes and transformation to standard format. Levenson et al. (2023) point out that the three-dimension metrics are built up prior due diligence and critical tasks on evaluating data curation, ascertaining and validating key variables using appropriate approaches as outlined earlier in this section.

## 3.3 Special considerations for different types of master protocols using EDE

In addition to the points discussed in Sections 3.1–3.2, we discuss in this section some special considerations for basket, umbrella, and platform trials with respect to design and analysis elements when using external controls.

- *Study populations.* Different patient cohorts of external control groups are usually required for a basket trial with each of the patient cohort corresponding to a single disease or disease subtype, while the same patient cohort of external control is often used for all treatment regimen in a umbrella trial. On the other hand, either different cohorts or the same cohort of external control patients or (internal) NCCs may be used in platform trials, depending the specific IPs and/or disease areas being investigated. This leads to extra complexity for defining and identifying external control patients, especially for basket trials. For example, multiple external data sources (e.g., disease registries) can be used in basket and platform trials and/or a single (or multiple) comprehensive large external data source containing pre-defined SOCs can be used to identify external control patients.

- *Adaptive features.* Adaptive strategies are commonly used in master protocols, e.g., sample size re-estimation in basket trials (Beckman et al., 2016) and adaptive randomization in umbrella trials (Kim et al., 2011) and platform trials (Berry et al., 2015). Adaptation often leads to changes in study population (e.g., biomarker or response adaptive) and treatment elements (e.g., schedule, cycle, dosage, and duration), which creates extra complexity in choosing external controls. For a biomarker-driven (or response-adaptive) basket trial with external controls, one may consider (1) whether the biomarker information can be captured in external (especially historical) data, (2) re-estimation of the size of external control patients for adaptive randomization or enrichment design, and (3) the composition of different disease subtypes in the external control if a common control is used. For umbrella and platform trials with an internal control, there is a high possibility that the internal control is non-concurrent and/or augmented with external controls, resulting in temporal variability in patient characteristics between NCC (or external control if augmentation is used) and treatment arms.

- *Outcome assessment.* Consistency in outcome assessment (e.g., method and scheduling) is one of critical components to ensure minimum bias in treatment effect estimation. Often different substudies use different methods (in basket or platform trials for different diseases or disease subtypes) or scheduling (in umbrella or platform trials for different IPs) for outcome assessment (Zanello1 et al., 2023), which can be very different from that in external controls. This discrepancy in outcome assessment between master protocol and external control can induce bias, e.g., disease assessment time bias (Bhattacharya et al., 2009).

- *Analytical perspectives.* For master protocols using external controls with interim analysis for efficacy, multiplicity adjustment should be considered if the objective focuses on individual indications, although bias in estimated effect outweighs type I error inflation. For umbrella and platform trials with NCCs that are augmented with external controls, the non-concurrent and external controls can be pooled if consistency is obtained using the test-then-pool approach (Viele et al., 2014); otherwise, Bayesian borrowing (Spiegelhalter et al., 2004) or synthetic methods discussed in Section 2.5 can be considered to form a single control arm.

## 3.4 Similarities and differences between MP trials and non-MP trials using external controls

Given the above discussion on master protocols using external controls and following the outline of the FDA guidance on considerations for the design and conduct of externally controlled trials (FDA, 2023), we provide a summary on the similarities and differences between master protocol and non-master protocol trials with respect to study design, external data sources and comparability assessment, analytical methods, and other considerations (e.g., diagnosis, follow-up period, and endpoint assessment) (Table 1).

## 4 A Targeted-Learning Roadmap for Causal Inference in Master Protocols using EDE

Clinical studies including master protocols using EDE for causal inference of treatment effects require careful considerations to ensure transparency of process steps, interpretability of causal analysis results, and robustness of study conclusion to causal assumptions. The ASA BIOP RWE SWG in its phase I project performs a landscape assessment on the use of causal inference frameworks for study design and analysis and propose a targeted learning (TL)-based causal inference roadmap (Ho et al., 2023) and in its phase II project illustrates how the TL-based causal inference roadmap can be applied to RWE studies (Ho et al., 2023). The TL-based causal inference roadmap, a systematic approach to answering causal questions, provides step-by-step guidance for transforming clinical study data into reliable evidence through precise description of research questions, definition of estimands, connection with observed data, specification of appropriate analysis, identification of causal assumptions necessary for causal interpretation of results, and assessing the impact of possible violation of causal assumptions on study conclusion. The TL-based roadmap, as one of the FDA's RWE demonstration projects (Concato, 2022), is also relevant to master protocols using EDE and can be summarized as follows. See also Petersen and van der Laan (2014), Gruber et al. (2023), and Ho et al. (2023) for more discussions on TL framework.

## 4.1 Step 0: Clearly defined research question

As discussed in Section 3.1, a master protocol often constitutes multiple substudies, each of which tries to answer a specific clinical question. For example, the MORPHEUS trial was designed to show whether the combination immunotherapy has anticancer activities for patient populations with selected specific cancer types (see Section 5.2). Each research question can be precisely described by the five attributes of estimands as outlined in the ICH E9(R1) guidance (ICH, 2021) and further discussed specifically for master protocols using EDE in Section 3.1. For example, the question of interest for a substudy in a master protocol can be "What is the effect of a new therapy relative to an SoC in patients with hepatocellular carcinoma who meet eligibility criteria and adhere their assigned therapy?", which is equivalent to the per-protocol (PP) analysis.

## 4.2 Step 1: Observed data and its generating mechanism

Following the principles and methods discussed in Section 3.2, one performs fit-for-use external data assessment and selects ones that can best address the clinical questions of interest as defined in Step 0. Upon choosing the fit-for-use external data, one describes the data-generating mechanism of the observed data. For a master protocol with an external control, the target population is usually defined by a set of inclusion and exclusion criteria in the master protocol and the criteria may not easily be used to select control patients from external database because of discrepancies in covariate measures (Hernán and Robins, 2016). Therefore, it is helpful if the target patient population in the master protocol can be simulated from the external data such that the two sets of patients (in master protocol and external control) are compatible.

More generally, the observed data structure $O$ for a one-time treatment study can be expressed in the form $O = (C, A, \Delta, \Delta Y)$, where $C$ represents the patient's baseline covariates, $A$ the treatment assigned to the patient, $\Delta$ the censoring status or adherence to assigned treatment (1 for adherence and 0 for non-adherence), and $Y$ the outcome of interest (e.g., 1 for tumor response and 0 for non-response). For longitudinal data with multiple decision points and time-varying covariates, e.g., a patient may have a dose adjustment or discontinue the assigned treatment at the $t$th time point (visit), which depends on the time-varying covariates (including disease status change), then the observed data $O$ can be written as $O = ((C_0, A_0, \Delta_1), (C_1, A_1, \Delta_2), \ldots, (C_t, A_t, \Delta_{t+1}), \ldots, (C_T, A_T, \Delta_{T+1}, \Delta_{T+1}Y))$; see van der Laan and Rose (2018), Ho et al. (2023), and Gruber et al. (2023) for more discussion on longitudinal data structure for targeted learning.

## 4.3 Step 2: Statistical model and targeted statistical estimand (parameter)

A statistical model specifies a collection of probability distributions of the data from both master protocol and external control with underlying plausible statistical assumptions. For a single-decision point in which, e.g., a rescue therapy or a dose adjustment may not be considered as part of the treatment regimes, then a regression model can be constructed, with or without interactions of treatment with prognostic and/or predictive factors. For longitudinal structured data, the TL method defines a statistical model by respecting the time ordering of data generating process. For example, the conditional distribution of the final outcome $Y$ is modeled based on $(C, A, \Delta)$ in which $\Delta$ is modeled conditionally based on $(C, A)$ and $A$ is modeled based on patient's baseline covariates $C$ (i.e., propensity score).

## 4.4 Step 3: Causal model and causal estimand

A causal model defines causal relationship among all relevant variables including covariates, treatment, and outcome in the observed data, based on our domain knowledge and causal assumptions. Domain knowledge includes disease etiology and pathophysiology, mechanism of action of the drugs used in the master protocol and external controls, prognos-tic and predictive factors and their relations to the treatment and outcome. The causal model can be written as a system of structural equations that express each endogenous (observed) variable as a function of its parents and an exogenous (unobserved) variable. For example, $C = f_c(U_c)$, $A = f_a(C, U_a)$, $\Delta = f_\delta(C, A, U_\delta)$, and $Y = f_y(C, A, U_y)$, where $O = (C, A, \Delta, \Delta Y)$ are endogenous variables and $U = (U_c, U_a, U_\delta, U_y)$, exogenous variables, and $f = (f_c, f_a, f_\delta, f_y)$ are deterministic functions describing the relation of each endogenous variable with its parents and an exogenous variable.

Upon specifying the system of structural equations, one can define a causal quantity (estimand) of interest. For example, the average treatment effect (ATE), defined as the difference in mean outcome that would have been observed had all subjects received the assigned treatment versus had all subjects received the control, can be expressed as $\psi_{\text{ATE}}^{\text{causal}} = E(Y_1) - E(Y_0)$, where $Y_1$ (or $Y_0$) denotes the counterfactual outcome a subject would experience when receiving treatment (or control). The expectation is taken with respect to the entire population included in both the substudies of master protocol trial and the external control.

## 4.5 Step 4: Statistical estimand versus causal identifiability

The statistical estimand is defined with respect to the observed data distribution $P_0$. For example, the statistical estimand $\psi^{\text{stat}}(P_0) = E_{c,0}[E_0(Y|A = 1, \Delta = 1, C) - E_0(Y|A = 0, \Delta = 1, C)]$ can be used to estimate the causal estimand $\psi_{\text{ATE}}^{\text{causal}}$ if identifiability assumption holds. A causal estimand is said to be identifiable from the observed data if the following assumptions are satisfied—Consistency, positivity, and exchangeability (Hernán and Robins, 2022). *Consistency* concerns whether the observed outcome for every treated (untreated) individual equals to the potential outcome should the person receive treatment (control). Hence, consistency connects the potential outcome with the observed data, i.e., $E(Y_a|A = a) = E(Y|A = a)$ for $a \in (0, 1)$. If consistency assumption holds, one observes $Y = Y_1$ when treatment $a = 1$ and $Y = Y_0$ when treatment $a = 0$; if consistency assumption does not hold, one cannot observe any potential outcome. *Positivity*, a.k.a. common support or overlapping, refers to as a non-zero (positive) probability of treatment assignment for all treatments and for all level of covariates. With positivity assumption, one can ensure that there is a probability greater than zero of assigning an individual to each of the treat-ment level. Positivity is required only for covariates *C* that are required for exchangeability. *Exchangeability* implies independence between the potential outcome and the actual treatment, i.e., $Y_a \perp A$ for all values $a$ which is called marginal exchangeability. Exchangeability can also be conditional when the potential outcome and the actual treatment are independent within a stratum of covariates $C = c$, i.e., $Y_a \perp A \mid C = c$. In general, exchangeability implies equal distributions of prognostic factors for the outcome between the treated and untreated in the entire sample or within a stratum. These causal assumptions are essential for bridging causal estimands with corresponding statistical estimands. See also Hernán and Hernández-Díaz (2012), Pearl et al. (2016), van der Laan and Rose (2018), and Ho et al. (2023) for more discussions on identifiability assumptions of causal estimands.

## 4.6 Step 5: Estimation of statistical estimand

Upon defining the statistical estimand, one can apply a pre-specified statistical method with a suitable estimator to estimate the target statistical estimand $\psi^{\text{stat}}(P_0)$. Petersen and van der Laan (2014) point out that different estimators may have different statistical properties that can result in meaningful differences in performance. For example, an inconsistent estimator can lead to statistical bias and its confidence interval may have a poor coverage of the target statistical estimand. Therefore, choice among estimators should be driven by their statistical properties and their underlying performance in a given subject matter. Some popular methods to estimate the target statistical estimand include propensity score method (Rosenbaum and Rubin, 1983), prognostic score method (Hansen, 2008; Stuart et al., 2013), parametric *G*-computation (Robins, 1986), inverse probability weighting (IPW) (Robins et al., 1994), targeted maximum likelihood estimation (TMLE) (van der Laan and Rubin, 2006; van der Laan and Rose, 2011), and machine learning methods (Guo et al., 2019; Kaddour et al., 2022).

For example, in *G*-computation method, one first fits the regression model of outcome *Y* on the treatment *A* and covariates *C* to obtain the conditional expectation $E(Y \mid A, C)$ and then uses the fitted model to predict the outcome for each subjects in the sample by setting the treatment to the level of interest $A = a$ and keeping the covariates unchanged, i.e., $\hat{E}(Y_i \mid A_i = a, C_i)$ for $a \in (0, 1)$ and $i = 1, \ldots, N$. The estimate for the ATE is obtained as a sample average of the difference in the predicted outcomes by adjusting the empirical distribution of $C_i$, i.e., $\hat{\psi}^{\text{G-comp}} = \sum_{C_i}[\hat{E}(Y_i|A_i = 1, C_i) - \hat{E}(Y_i|A_i = 0, C_i)]P(C_i)$. As another example, in the TMLE method, the probability distribution of the observed data $O$ can be factorized as $P(O) = P(Y|A, C)P(A|C)P(C)$. Define $Q(Y, A, C) = E(Y|A, C)$ and $g(A, C) = P(A|C)$, where $Q(\cdot)$ is estimated from data $O$, $g(\cdot)$ can be further factorized into treatment, missingness, and censoring mechanisms (using appropriately defined function of $\Delta$), and the empirical distribution $P(C)$ of $C$. The TMLE estimate for ATE is $\hat{\psi}^{\text{TMLE}} = \frac{1}{n}\sum_{i=1}^{n}[Q_n^*(1, c_i) - Q_n^*(0, c_i)]$. To obtain $Q_n^*(a, c_i)$, one first fits a model to estimate the conditional mean outcome $Q_n^0(a, c_i) = \hat{E}^0(Y|A, C)$ given $(A = a, C = c)$ and a model $g(1, c)$ for the propensity score $P(A = 1|C = c)$. With $Q_n^0$ and $g(a, c)$, one aims to improve the estimator of the targeted parameter of the distribution by obtaining the MLE for $\epsilon$ in $Q_n^1 = Q_n^0 + \epsilon h(A, C)$, where $h(A, C)$ is a function of the nuisance parameter that depends on the influence curve of the parameter of interest. See van der Laan and Rose (2011) and Gruber and van der Laan (2012) for more details.

## 4.7 Step 6: Result interpretation and sensitivity analysis

Petersen and van der Laan (2014) provide a hierarchy of interpretation with increasing strength of assumptions necessary for statistical results to be interpreted causally. For example, the estimate $\hat{\psi}^{\text{stat}}$ and its associated variance can be interpreted as the difference and uncertainty in mean outcome between the treated and untreated subjects, averaged over the distribution of measured covariates in the target population. For the target statistical estimand $\psi^{\text{stat}}$ to have a causal interpretation, identifiability assumptions are required to ensure that the causal estimand is identifiable from the observed data, i.e., the chosen causal model accurately describes the true generating mechanism of the observed data. Since these assumptions are mostly unverifiable, its plausibility can be explored through a sensitivity analysis for which an essential step is to pinpoint which assumption(s) could be violated and under what degree. For example, in a master protocol trial with an external control, exchangeability assumption is likely violated because no randomization is used for treatment assignment; positivity assumption may be violated if there are only the treated (or untreated) subjects in some strata of covariates in the study sample; and consistency may also be violated if outcomes are mis-measured or measured inconsistently between subjects in the master protocol and those in the external control.

Causal bias occurs when any of the causal assumptions are violated, leading to the causal gap $\eta = \psi^{\text{causal}} - \psi^{\text{stat}}$ which can be assessed by a sensitivity analysis through a variety of ways. For example, one can perform a sensitivity analysis by (1) assessing the impact of various degrees of violation to the identifiability assumptions on the implications of the statistical results for the causal estimand, (2) investigating how the effect estimate and its confidence bounds change over a range of plausible values of the causal gap $\eta$, (3) using the target trial framework of Hernán and Robins (2016) to emulate an RCT for causal inference of treatment effects, or (4) decomposing the bias into two main components—statistical bias and causal bias—and using the targeted learning framework to reduce the statistical bias and constructing the bounds on causal bias (Díaz et al., 2018). Using the targeted learning framework, Gruber et al. (2023) illustrate the impact of the magnitude and direction of the hypothetical causal gap $\eta$ on the effect estimate and its confidence bounds and Ho et al. (2023) demonstrate how a sensitivity analysis helps ascertain whether the substantive conclusions continue to hold or break down under plausible levels of a causal gap $\eta$. Lash et al. (2014) summarize good practice for quantitative bias analysis, which is equivalent to the 4-step approach of Díaz et al. (2018), and point out that modeling of bias needs to be performed in reverse order from what they occur. For example, confounding bias in source population occurs before selection bias in selecting subjects which takes place before information bias when measuring exposure, outcome, and covariates. VanderWeele and Ding (2017) propose the *E*-value as a tool to measure "the minimum strength of association, on the risk ratio scale, that an unmeasured confounder would need to have with both the treatment and the outcome to fully explain away a specific treatment-outcome association, conditional on the measured covariates."

In summary, the targeted learning-based causal roadmap described above provides a systematic approach to integrating causal modeling with statistical estimation. The principles of the roadmap are essentially the same as those in the estimand framework ICH (2021) and hence are applicable to both RCTs and RWE studies (including master protocols using external controls). The principles can be summarized in three key steps: (1) define a target statistical estimand that aligns with the causal estimand for the study objective, (2) use an efficient estimator to estimate the target statistical estimand and its uncertainty, and (3) evaluate the impact of causal assumptions on the study conclusion by performing a sensitivity analysis (Gruber et al., 2023; Ho et al., 2023).

## 5 Case Studies

### 5.1 MASTER KEY project

The MASTER KEY project, a platform trial embedded in a registry database (Okuma et al., 2020), consists of two parts: A prospective registry study and multiple clinical trials (substudies). The purposes of the registry study are (1) to build a highly reliable database that contains patient backgrounds, biomarker information, and prognostic variables to help understand the nature of patients with rare cancers; (2) to establish reliable historical control data for clinical trials; and (3) to assign enrolled patients to clinical trials based on their molecular profiles, i.e., patients in the registry study may receive either a biomarker-directed treatment inside the MASTER KEY clinical trials or routine clinical practice. The clinical trial part of the MASTER KEY includes multiple substudies and aims at obtaining regulatory approval and national health reimbursement under the "conditional early approval system" in Japan. Each substudy will be based on a single-arm study with the primary end point of response rate. Patients who receive routine clinical practice are followed in the registry as part of the external control data. The five attributes of the primary estimand for the MASTER KEY project can be described as follows: (1) *Population*: (a) Age 1 year or older, (b) histological diagnosis of rare cancer, cancer of unknown primary origin, rare tissue subtype of common cancer, or hematologic malignancy, (c) incurable progressive disease, and (d) certain molecular profiles; (2) *Treatment*: Pharmacotherapy, radiotherapy, chemotherapy; (3) *Endpoint*: Objective response to assigned treatments; (4) *Intercurrent events*: Treatment failure; and (5) *Population-level summary*: Objective response rate.

The authors state that patients would be given other options after treatment failure which presumably include both "fail to tolerate" and "fail to show efficacy" (lack of efficacy). In addition, patients may experience dose-adjustment or terminal events (e.g., death), or self-select preferred therapies during the trial, which should be clearly stated on how these ICEs would be accounted when estimating the primary estimand. In "trial design and statistical analysis" section, the authors specify the null hypothesis $H_0 : p \leq p_0$ and the alternative hypothesis $H_1 : p \geq p_1$ for response probability (RP) $p$, where $p_0$ denotes the lower futile RP and $p_1$ the desired minimum clinically meaningful RP. They state that "the target and desired response rates are individually determined for every substudy based on historical information on the effect of conventional treatment for the cancer type we test in the trial." However, the MASTER KEY project also includes patients in routine clinical practice who serve as "contemporaneous control" patients. Okuma et al. (2020) provide a comparison of progression-free survival between patients who participate in clinical trials and those who receive routine clinical practice (Figure 3 (b)) and between patients who receive biomarker-directed treatment and those who don't (Figure 3 (c). The authors acknowledge that the comparison may suffer from confounding by prognostic and predictive biomarkers, patient selection bias, and bias due to nonconcurrent historical controls. Perhaps a more scientifically rigorous approach is to (1) use the estimand framework discussed in Section 3.1 to precisely define the estimand in each substudy by focusing on the comparability of study population with respect to baseline and time-varying covariates and other potential intercurrent events (besides treatment failure) and the pattern of their occurrence between treated patients and control patients, (2) apply the target-learning roadmap presented in Section 4 to connect the causal estimand with the observed data (including the historical controls) by explicitly specifying causal assumptions, and (3) perform sensitivity analyses by investigating the impact of potential violation of causal assumptions, either individually or in combination, on the study conclusion.

## 5.2 MORPHEUS

MORPHEUS is a phase Ib/II platform study consisting of multiple, global, open-label, randomized, phase Ib/II trials to evaluate multiple investigational drugs in patients with different tumor types (Desai et al., 2019; Helms, 2020; Li et al., 2022). It is designed to identify early signals of clinical activity of combination cancer immunotherapies (CITs) and establish proof-of-concept that supports the cross-indication learning with transformational potential. In MORPHEUS platform, treatment arms showing minimal clinical activity or unacceptable toxicity will be closed, while new treatment arms can be open as novel combinations become available. By November 2022, nine different trials were registered at the clinicaltrials.gov, with each focusing on one tumor type including metastatic colorectal cancer (MORPHUES-CRC). In the MORPHEUS-CRC trial, patients with mi-crosatellite stable tumors who had been refractory to the firs and second lines of therapies were randomized to either one of eight experimental arms or a control arm with regorafenib, a standard-of-care therapy in the disease setting (clinicaltrials.gov, NCT03555149). Li et al. (2022) report the primary results of the experimental arm (atezolizumab + isatuximab) and the control arm (regorafenib) with the hybrid control design in which data of eligible patients from the external control arm of a completed clinical trials (IMblaze370, NCT02788279) were incorporated into the MORPHEUS-CRC concurrent control data to construct a hybrid control arm using a frequentist model with PS weighting or a Bayesian dynamic borrowing. See Figure 2 for an overview of the study design and analysis methods.

The study by Li et al. (2022) can be summarized using the estimand framework as follows: (1) *Population*: (a) patients with microsatellite stable tumors who had been refractory to the first and second line of therapies, (b) measurable disease per RECIST v1.1; (c) tumor accessible for biopsy; (2) *Treatment* : eight CIT combinations versus a control (regorafenib); (3) *Endpoint*: Investigator-assessed objective response as the primary endpoint for MORPHEUS-CRC and disease-control rate, investigator-assessed progression-free survival and overall survival as the key secondary endpoints in the hybrid control analysis; (4) *Intercurrent events*: Treatment switches; (5) *Population-level summary* : Overall response rate. Li et al. (2022) describe methods used to MORPHEUS-CRC experimental arm and the external control arm in terms of baseline covariates.

In additional to the analytical methods discussed in Li et al. (2022) to balance baseline covariates between experimental arm and the hybrid control arm, some additional considerations may help further elucidate the causal estimate of treatment effects. First, it is unclear how ICEs (e.g., treatment discontinuation due to lack of efficacy or intolerability) are defined and what strategies are taken to address them. For example, a patient may use a rescue therapy when disease progresses or switches to another therapy when the patient cannot tolerate the assigned treatment. This can be done by explicitly specifying ICEs in Step 0 to reflect the research question. Second, although both MORPHEUS-CRC and IMblaze370 trials use investigator-assessed tumor response with the former following up the patients up to approximately 3–5 years and the latter up to approximately 20 months, it is unclear whether the tumor assessment schedules are approximately equal between the two trials. Bhattacharya et al. (2009) point out that different tumor assessment patterns between arms may cause tumor assessment time bias. For example, a shorter time interval between scheduled visits generally results in an earlier tumor progression. Third, bias may be induced by different tumor assessment methods and non-radiologic (clinical) assessment is often subjective. The potential difference in assessment time and methods can be modeled in Steps 2–4 to investigate whether different assessment intervals and methods generate bias of treatment effect and if yes, the size and direction of bias. Fourth, both trials are open-label studies, which may lead to *treatment heterogeneity bias*, i.e., treatment effect is correlated with treatment status (e.g., self-selection propensity for a particular treatment) (Vaupel and Yashin, 1985; Xie, 2011, 2013), which is less recognized in current practice. Finally, explicit discussions on causal assumptions and a sensitivity analysis to explore the impact of potential violation of these assumptions may help causally interpret the results and study conclusion. Again, this can be done in Steps 3 and 6 for either individual assumptions or multiple assumptions simultaneously.

In additional to the analytical methods discussed in Li et al. (2022) to balance baseline covariates between experimental arm and the hybrid control arm, some additional considerations may help further elucidate the causal estimate of treatment effects. First, it is unclear how ICEs (e.g., treatment discontinuation due to lack of efficacy or intolerability) are handled and what strategies are taken to address them. For example, a patient may use a rescue therapy when disease progresses or switches to another therapy when the patient cannot tolerate the assigned treatment. Second, although both MORPHEUS-CRC and IMblaze370 trials use investigator-assessed tumor response with the former following up the patients up to approximately 3–5 years and the latter up to approximately 20 months, it is unclear whether the tumor assessment schedules are approximately equal between the two trials. Bhattacharya et al. (2009) point out that different tumor assessment patterns between arms may cause tumor assessment time bias. For example, a shorter time interval between scheduled visits generally results in an earlier tumor progression. Third, bias may be induced by different tumor assessment methods and non-radiologic (clinical) assessment is often subjective. Fourth, both trials are open-label studies, which may lead to *treatment heterogeneity bias*, i.e., treatment effect is correlated with treatment status (e.g., self-selection propensity for a particular treatment) (Vaupel and Yashin, 1985; Xie, 2011, 2013), which is less recognized in current practice. Finally, explicit discussions on causal assumptions and a sensitivity analysis to explore the impact of potential violation of these assumptions may help causally interpret the results and study conclusion.

## 6 Discussion and Conclusion

There is massive literature discussing external (including historical) controls in clinical trials when RCTs are infeasible or unethical. This paper has provided a consolidated overview about different types of external controls and some specific considerations when using them in master protocols. When fit-for-use high-quality external data are available, contemporaneous controls are in general preferred over historical-contemporaneous controls which are favored over historical controls because most data elements (e.g., data points to be collected by what methods) can be prospectively defined and the external data are collected simultaneously with the data in the master protocol. In addition, data from prior clinical trials are generally believed more reliable; however, their relevance needs to be carefully reviewed to avoid confounding bias and selection bias (Mishra-Kalyani et al., 2022). Note that, depending on the objectives and study design, a master protocol investigating multiple products on the same disease may use a common external control and a master protocol investigating multiple disease (sub)types may use separate external controls derived from specific data sources (e.g., disease registries), each of which would best possibly serve its purpose for each substudy in the master protocol.

Chen et al. (2023) present the scientific, regulatory, ethical, and operational considerations as the rationale for using external data in the design and analysis of clinical trials, which is also applicable to master protocols. Although outcome variables are a critical data element in assessing fit-for-use external data, selection of patients from the fit-for-use data to compose the external control arm should only be based on prospectively defined covariates, not on the observed outcome variables (Yue et al., 2014). In addition, sample size determination for the external controls should take into account the analysis methods for treatment effect estimation and should be adaptive, i.e., preliminary estimation in the design stage and final confirmation upon regulatory agreement to key aspects (e.g., selection of key covariates to create comparable groups, analytic methods for modeling treatment effects, and methods for type I error control).

It is perceivably understood that all estimands should be pre-defined before a study be-gins. However, considering adaptive natures of master protocols, there are scenarios where an estimand can only be clearly defined until after a substudy starts. For example, adding a new treatment arm to a platform trial simply implies adding another objective and hence another estimand to the protocol, with or without use of external controls. Occasionally, external evidence may become available during a master protocol, which may suggest modification of certain aspects of the trial, such as sample size reassessment, discontinuation of a treatment arm, or an unplanned change of primary endpoint, etc., and hence likely the estimand change in an ongoing study Collignon et al. (2022).

The estimated statistical parameters (statistical estimands) may not be robust and reliable due to small sample sizes of substudies in master protocols. To overcome this challenge, one may consider different statistical methods (e.g., Bayesian methods) for parameter estimation. Or machine learning and super learning approaches (van der Laan et al., 2007; Polley and Van Der Laan, 2010) (as briefly mentioned in Section 2.7) can be used to achieving optimal results for semi-parametric estimation of causal effects such as in Colson et al. (2016). Alternatively, one can construct predictive models by applying machine learning and super learning methods to external control data (or experimental arm data) and then use these models to predict potential outcomes for patients in the master protocol (or external control) such as in Loiseau et al. (2022).

In summary, it is highly recommended that a master protocol using EDE follows the target-learning roadmap starting from articulating the research question that determines the causal estimand, specifying the statistical model and the target statistical estimand, assessing causal identifiability, estimating the target statistical estimand, and performing sensitivity analyses to evaluate robustness of analysis results to study conclusion.

## Acknowledgements

We would like to thank ......## References

Abadie, A. (2021). Using Synthetic Controls: Feasibility, Data Requirements, and Methodological Aspects. *Journal of Economic Literature 59* (2), 391–425.29

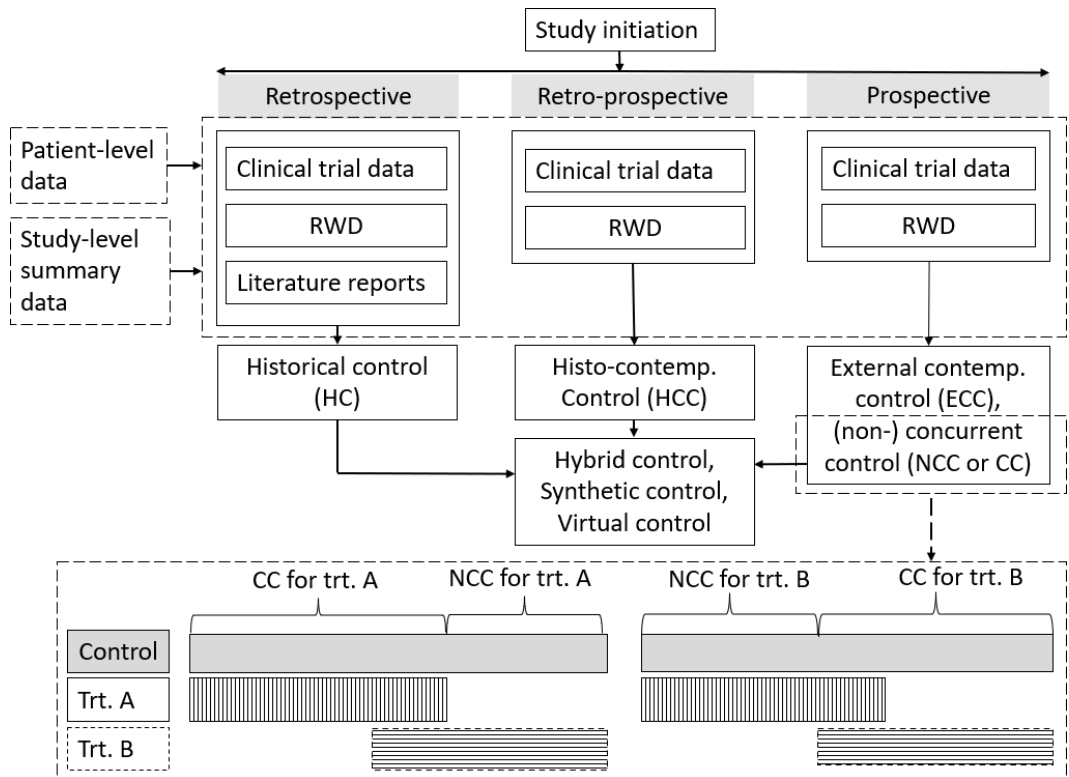

Figure 1: Types and data sources of controls commonly used in master protocols (Contemp.=Contemporaneous; Histo-contemp.=Historical-contemporaneous; trt=treatment). The bottom dashed rectangle shows (internal) concurrent control (CC) and non-concurrent control (NCC) for either treatment A or treatment B; see Sridhara et al. (2022) for further discussion on concurrent and non-concurrent controls.

Table 1: Similarities and differences in using external controls between non-master protocols and master protocols.

| Similarities/ differences | Non-Master Protocols | Master Protocols |
|---|---|---|
| | *Design considerations* | |
| Similarities | • Pre-specification of the following critical elements in the protocol regarding the external control: suitable data sources and their fitness assessment, subject eligibility criteria, exposure definitions and windows, well-defined endpoints, appropriate analytic methods, missing data handling | |
| Differences | • A single estimand for each objective, e.g., primary estimand for primary objective<br>• A single definition for each element of an estimand for the whole study<br>• A universal baseline and treatment (index) window for all subjects in the external control<br>• The same set of potential confounding variables for all subjects in the study<br>• Methods handling confounding variables applied to all subjects in the study | • Multiple estimands with each for a substudy within a master protocol<br>• Precise definition of the elements of an estimand for each substudy<br>• Perhaps different baseline and treatment (index) windows for different substudies depending on the disease and SoC for the disease<br>• The sets of potential confounding variables could differ between substudies<br>• Different methods for confounding adjustment to accommodate different confounding situations in the studies |
| | *External data sources and comparability assessment* | |
| Similarities | • Data sources may include completed clinical trials and RWD<br>• Comprehensive assessment of fit-for-use external data with a focus on measurement of baseline covariates defining target population, treatments, and outcomes (endpoints) of interest<br>• Comparability assessment may include both baseline and time-varying covariates | |

Table 1 – continued from previous page

| Similarities/ differences | Non-Master Protocols | Master Protocols |
|---|---|---|
| Differences | • A cohort from a single data source is usually selected (occasionally multi-data sources are used for rare diseases)<br>• Data comparability is assessed at the study level with a focus on unified definitions of key covariates (e.g., treatment regimes, diagnosis, prognosis, follow-up period, intercurrent events, and outcome) | • Multiple cohorts from either a single data source or multiple data sources are used to form external control arms, each of which corresponding to a substudy<br>• Data comparability is assessed at the substudy level with more complex definitions of key covariates, which are uniquely defined for individual substudies |
| *Analytical methods* | | |
| Similarities | • Statistical analysis plan should be pre-specified along with study protocol<br>• Pre-defined analytical methods to be used for confounding adjustment<br>• Analytical methods may include matching, stratification, regression (e.g., *G*-methods), doubly robust methods (e.g., TMLE), and machine learning methods | |
| Differences | • A singe set of pre-defined analytical methods for primary analysis is usually applied to all subjects in the study<br>• A single set of sensitivity analysis is used to evaluate the robustness of results to causal assumptions | • Different analytical methods may be applied to accommodate possibly different causal assumptions between substudies<br>• Multiple sets of sensitivity analysis methods may be applied to evaluate the impact of causal assumptions on study results |
| *Other considerations* | | |
| Differences | • A universally acceptable diagnostic method is usually pre-defined for all subjects<br>• The same duration of follow-up period is applied to all subjects<br>• Methods for endpoint assessment are the same or similar for all subjects | • Different diagnostic methods are generally used in different substudies<br>• Different durations of follow-up periods may be applied in different substudies<br>• Methods for endpoint assessment may differ in different substudies |

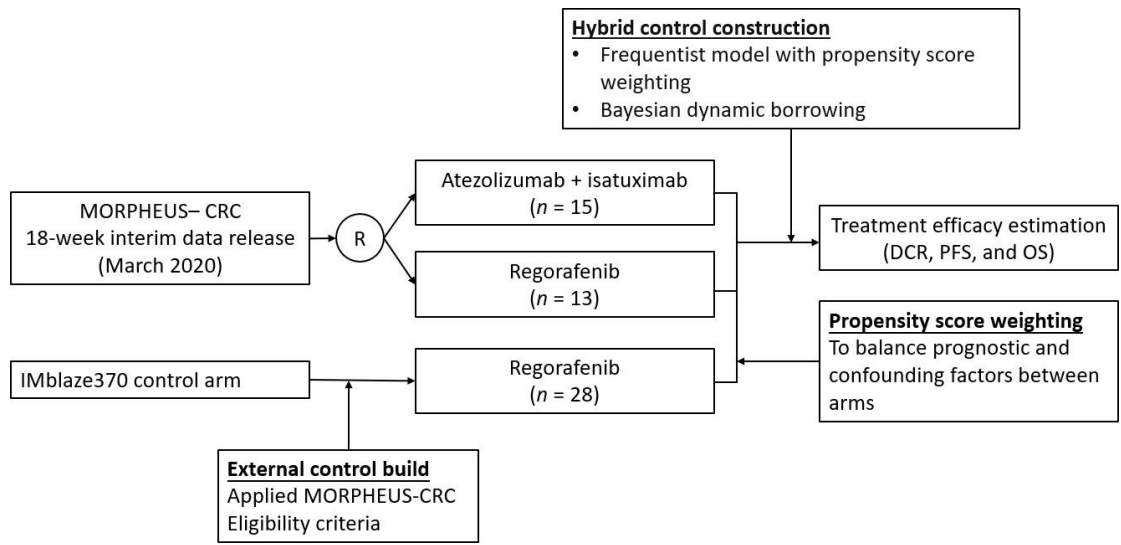

Figure 2: An overview of the MORPHEUS-CRC design and analysis. DCR–disease control rate, PFS–progression-free survival, OS–overall survival (Li et al., 2022).